\documentclass[conference]{IEEEtran}
\IEEEoverridecommandlockouts
\usepackage{cite}
\usepackage{amsmath,amssymb,amsfonts}
\usepackage{algorithmic}
\usepackage{graphicx}
\usepackage{textcomp}
\usepackage{xcolor}
\usepackage{url}
\usepackage{hyperref}
\usepackage{booktabs}
\usepackage{multirow}
\usepackage{caption}
\def\BibTeX{{\rm B\kern-.05em{\sc i\kern-.025em b}\kern-.08em
    T\kern-.1667em\lower.7ex\hbox{E}\kern-.125emX}}

\begin{document}

\title{TTS-Transducer: End-to-End Speech Synthesis with Neural Transducer}

\author{
\IEEEauthorblockN{
    Vladimir Bataev\textsuperscript{*}, Subhankar Ghosh\textsuperscript{*}, Vitaly Lavrukhin, Jason Li
}
\IEEEauthorblockA{
\textit{NVIDIA} \\
\{vbataev, subhankarg, vlavrukhin, jasoli\}@nvidia.com}
}

\maketitle
\begingroup\renewcommand\thefootnote{*}
\footnotetext{Equal contribution}
\endgroup

\begin{abstract}
This work introduces TTS-Transducer -- a novel architecture for text-to-speech, leveraging the strengths of audio codec models and neural transducers. Transducers, renowned for their superior quality and robustness in speech recognition, are employed to learn monotonic alignments and allow for avoiding using explicit duration predictors. Neural audio codecs efficiently compress audio into discrete codes, revealing the possibility of applying text modeling approaches to speech generation. However, the complexity of predicting multiple tokens per frame from several codebooks, as necessitated by audio codec models with residual quantizers, poses a significant challenge. 
The proposed system first uses a transducer architecture to learn monotonic alignments between tokenized text and speech codec tokens for the first codebook. Next, a non-autoregressive Transformer predicts the remaining codes using the alignment extracted from transducer loss. The proposed system is trained end-to-end. 
We show that TTS-Transducer is a competitive and robust alternative to contemporary TTS systems\footnote{Audio Examples are available at \url{https://tts-transducer.github.io}}.
\end{abstract}


\section{Introduction}
\label{sec:intro}
Neural text-to-speech (TTS) is a sequence-to-sequence task where the model learns to generate a speech sequence conditioned on the input text sequence. 
TTS synthesis is monotonic, preserving the order between input text and output speech. Since speech is produced at the frame level, one phoneme can correspond to multiple frames, and output length varies with the speaker's style, making text-to-speech alignment challenging.
Non-autoregressive (NAR) TTS models~\cite{ren2019fastspeech,beliaev2021talknet} use explicit phoneme or text token duration predictor. In autoregressive (AR) encoder-decoder~\cite{sutskever2014} TTS models~\cite{shen2017tacotron2}, alignment is learned implicitly, while they produce more natural speech, they also suffer from hallucination, skipping or repeating words~\cite{ping2018deepvoice3}.

The transducer architecture (RNNT)~\cite{graves2012transducer}, widely used in automatic speech recognition (ASR), enforces a monotonic alignment constraint. Thus, it could provide a robust solution for this problem. However, direct application of transducers to TTS is challenging since transducers are designed to predict discrete units, but speech is typically represented in continuous form, e.g., with a mel-spectrogram. Recent development in neural audio codecs~\cite{defossez2022encodec,kumar2023daccodec,langman2024spectralnemocodec} allows to transform the audio prediction task into a discrete units prediction task. This approach significantly simplifies TTS pipeline, and many recent state-of-the-art TTS models follow this approach, e.g., VALL-E~\cite{wang2023valle}, T5-TTS~\cite{t5tts}, Bark~\cite{BarkTTS}, SpeechX~\cite{wang2023speechx}. Using discrete speech codes as a target makes the transducer a natural fit for TTS alignment.
However, audio codecs produce \textit{multiple codes per frame}, and directly applying RNNT to predict all the codes sequentially requires a huge amount of memory to compute the loss, since memory complexity depends on the product of the input and target sequence lengths. We address this challenge by introducing a two-component architecture that is trained end-to-end. \textit{The transducer component} predict the first codebook codes. \textit{A residual codebook head (RCH)} part iteratively predicts the remaining codes using the aligned encoder output and the predicted previous codebook codes.
The components are conditioned on a speaker embedding, generated from the target speaker's sample speech using Global Style Tokens (GST)~\cite{wang2018globalstyletoken}. The model is based on the Transformer~\cite{vaswani2017attentionTransformer} architecture.

The key contributions of our work are as follows:
\begin{enumerate}
    \item A novel end-to-end TTS-Transducer model based on the neural transducer architecture that predicts audio codes directly from the tokenized text, solving the problem of producing multiple codes per frame.
    \item TTS-Transducer achieves $3.94\%$ character error rate (CER) on challenging texts, surpassing larger state-of-the-art TTS models trained on significantly more data.
    \item We demonstrate that TTS-Transducer is codec agnostic and is able to generalize well across different residual vector quantization (RVQ) codecs, which is are a rapidly growing field. 
\end{enumerate}
TTS-Transducer generates high-quality speech and achieves zero-shot results comparable to state-of-the-art TTS models without pretraining on large data. We will open-source our code in the NeMo toolkit~\cite{kuchaiev2019nemo}.

\section{Background and Related Work}

\subsection{Neural Transducers}
\label{sec:Transducers}

Recurrent neural network transducer~\cite{graves2012transducer} is a universal architecture for sequence-to-sequence tasks requiring the monotonic alignment between input and output. Transducer consists of three neural modules: (1)~\textit{encoder}, which processes the input sequence (originally audio features) and generates high-level representations, (2)~\textit{prediction network (predictor)} - an autoregressive network (originally RNN) that uses previously predicted tokens to generate the next output, starting from special  $\langle SOS \rangle$ (start-of-sequence) token, and (3)~\textit{joint network}, which combines outputs of encoder step $i$ and prediction network step $j$ to produce the distribution of the probabilities ($p_{i,j}$) over the vocabulary augmented with special $\langle blank \rangle$ ($\langle b \rangle$) symbol.
RNNT optimizes the total probability of all possible alignments between input and output sequences, where the target sequence includes inserted $\langle blank \rangle$ symbols, acting as delimiters between frames.

There are several decoding strategies for transducers. 
Greedy decoding uses nested loop design, where the outer loop iterates over frames of the encoder output, and the inner loop retrieves labels one by one with the maximum probability by combining the encoder output for the current frame and prediction network output using joint network until the $\langle blank \rangle$ symbol is found.
In our experiments, we apply nucleus sampling~\cite{holtzman2019nucleussampling} using predicted probabilities instead of greedy selection.

\subsection{Neural Transducers for Text-to-Speech Synthesis} 

Segment-to-Segment Neural Transduction (SSNT)~\cite{yasuda2019initial} introduced transducer-based monotonic restrictions in TTS but factorized joint probability into alignment transition probability and emission probability for acoustic features.
Speech-T model~\cite{chen2021speecht} also decouples aligning and mel-spectrogram prediction. The authors use a modified RNNT loss to learn the alignment, but the model requires the external forced aligner to construct diagonal constraints in the probability lattice to help the transducer learn the alignment. 

Recently introduced \textit{Transduce-and-Speak}~\cite{kim2023transduceandspeak} model has two components trained separately. The first component is a neural transducer generating  "semantic" tokens (one token per frame). Such tokens are obtained from the clustered output of pretrained Wav2Vec 2.0~\cite{Baevski2020wav2vec2}. The second component is a modified non-autoregressive (NAR) VITS~\cite{kim2021vits} model synthesizing speech from the semantic tokens.
A similar two-stage approach with intermediate "semantic" tokens was used in ~\cite{lee2024highfidelitytts} with the second component predicting audio codes.

More recently, the VALL-T~\cite{du2024vallt} combined transducer with a decoder-only Transformer. The authors combined relative position embeddings with absolute positional embeddings, where a relative position of 0 specifies the current phoneme under synthesis. The $\langle blank \rangle$ symbol in the output indicates a position shift. During training, the model uses all possible shifts of the positional embeddings, and the output is combined to form the transducer lattice. Transducer loss is used to optimize the model. VALL-T needs a large amount of memory and multiple forward passes during training due to all the relative position shifts.

\subsection{TTS using Audio Codes}

Recently, a new TTS approach has emerged, where TTS is considered a language modeling task that translates text input to discrete audio tokens. Similar to large language models (LLMs), there are two main types of models: (1) decoder-only (AudioLM, VALL-E, UniAudio, Bark, SpeechX~\cite{borsos2023audiolm, wang2023valle, yang2023uniaudio, BarkTTS, wang2023speechx}) and (2) encoder-decoder (e.g. Speech-T5 \cite{ao2021speecht5}).  
Typically audio codecs use an encoder-decoder architecture, e.g. EnCodec, SoundStream, Descript Audio Codec (DAC)~\cite{defossez2022encodec, zeghidour2021soundstream, kumar2023daccodec}. The encoder converts audio to a hidden representation, which the quantization layer compresses using the RVQ approach. The decoder synthesizes audio in the time domain from the quantized representation. The entire model is trained end-to-end using reconstruction loss and perceptual loss from a discriminator. 

\section{TTS-Transducer Model}
\label{sec:proposed-model}

\begin{figure}[t]
\centerline{\includegraphics[width=0.9\linewidth]{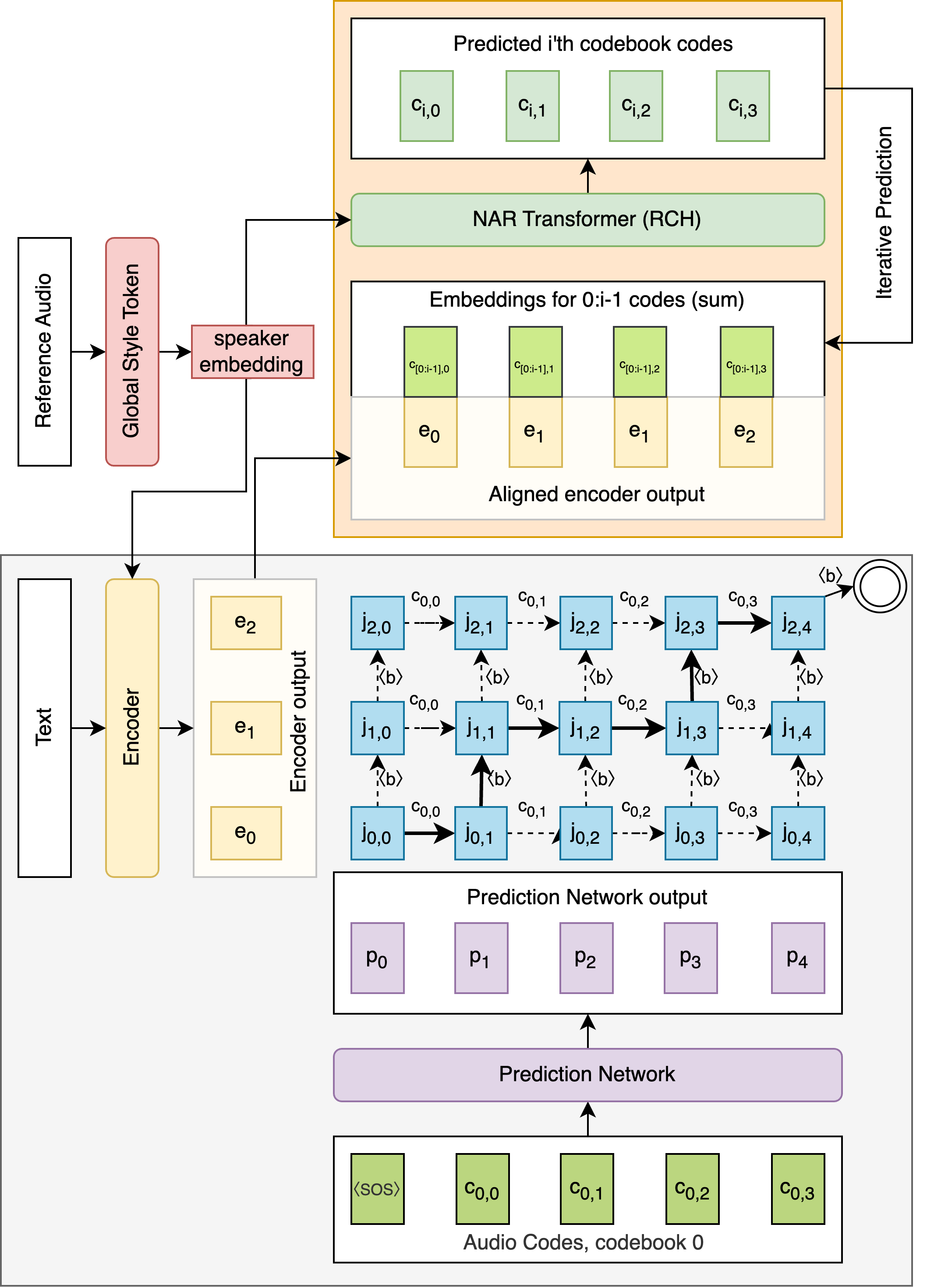}}
\caption{TTS-Transducer model architecture}
\label{fig:tts-Transducer}
\vspace{-15pt}
\end{figure}

The architecture of the TTS-Transducer, inspired by the VALL-E~\cite{wang2023valle} system, consists of two components trained separately. The first component of VALL-E is an autoregressive Transformer model that predicts codes of the first codebook given the input text and a prompt. The second one is a non-autoregressive Transformer that predicts codes from all other residual codebooks based on the prediction of the first component, input text, and a prompt.

TTS-Transducer schema is shown in Fig.~\ref{fig:tts-Transducer}. We use a neural transducer to predict the codes of the first codebook given the text units by learning alignment between text and audio. The second component, residual codebook head (RCH), is a non-autoregressive Transformer. It predicts the remaining audio codes iteratively, given all previously predicted codes along with the aligned encoder output. The encoder of the transducer and the residual codebook head are conditioned on speaker embeddings. After predicting all the codes, the decoder of the audio codec model is used to produce audio.

\textit{Prediction of the first codebook} $c_{0,i}$ is learned by a neural transducer. 
Encoder is a non-autoregressive Transformer~\cite{vaswani2017attentionTransformer} model, which transforms tokenized text $t_{i}$ to the sequence of vectors $e_{i}$. We train models with Byte-Pair Encoding (BPE)~\cite{bpe} tokenization, and also experiment with phonemes from International Phonetic Alphabet (IPA). 
% TODO: IPA tokenizer citation
We also add speaker embedding conditioning to the encoder using conditional LayerNorm~\cite{ba2016layernorm}. The prediction network is an autoregressive Transformer-Decoder, which transforms a sequence of audio codes $c_{0,i}$ with prepended $\langle SOS \rangle$ symbol to the sequence of vectors $p_j$. For each combination of vectors $e_i, p_j$ the joint network is applied: $j_{i,j} = Softmax(Linear(ReLU(e_i + p_j)))$. The output of the joint network is the probability distribution for the tokens of the first codebook augmented with the $\langle blank \rangle$ symbol.

\textit{To predict all residual codebooks} from $1$ to $n$, we use a non-autoregressive Transformer-encoder, which predicts $i$-th codebook codes using previously predicted $[0...i-1]$ codebooks and the aligned encoder output. The input is the sum of embeddings for previously predicted codes $c_{0:i-1,j}$ concatenated with the corresponding encoder vector $e_k$. We use speaker conditioning similar to the first component of our system. 

\textit{To represent speakers}, TTS systems typically use fixed embeddings from a speaker verification model~\cite{NEURIPS2018_6832a7b2}, but these embeddings do not generalize well beyond seen speakers. So we use Global Style Tokens (GST)~\cite{wang2018globalstyletoken} to capture the style of the speaker as used in ~\cite{hsieh23_interspeech}. 
In our work, we convert target speaker's reference speech to mel-spectrogram and feed it to the speaker representation module. The speaker representation module consists of a convolutional recurrent neural network-based encoder that learns the style tokens. A multi-head attention layer combines the learned style tokens to give the speaker embedding.

\textit{We train our system end-to-end.} On each training step, we first perform a forward pass for the first component of our system. We use WFST-based implementation of the RNNT loss~\cite{laptev2023rnntwfst} in the k2 framework~\cite{povey_k2}. This allows us to extract the alignment between audio codes and encoder output (corresponding to text units) from the calculated RNNT lattice~\footnote{To extract the alignment, we need to find the best path with maximum probability, which can be done with \texttt{k2.shortest\_path}.}. We distribute the encoder frames according to the extracted alignment. We also randomly select $i$ from all residual codebooks $[1:n]$ to predict $i+1$ codebook codes with the second part of our system.
We optimize this component by applying cross-entropy loss ($\lambda_{CE}$). The total loss is the weighted sum of the losses from the first and second components of our network: $\lambda_{total} = (1 - \alpha) * \lambda_{RNNT} + \alpha * \lambda_{CE}$. We use $\alpha = 0.4$ in our experiments.

\textit{In decoding}, we first evaluate the RNNT component, getting the predictions for the first codebook. Due to the nature of the decoding algorithm as described in Section~\ref{sec:Transducers}, getting the alignment along with the predictions does not imply computational overhead. For efficient decoding, we adopt label-looping~\cite{LabelLooping} greedy decoding algorithm, replacing greedy label selection with nucleus sampling~\cite{holtzman2019nucleussampling} on each step. Since our prediction network is a non-autoregressive Transformer, we use a key-value cache to speed up the decoding. After this step, for the remaining codebooks from $i$ to $n$, we iteratively evaluate our system's second component, utilizing the aligned encoder output. It is worth noting that the system can be naturally used in streaming mode by using an autoregressive Transformer as the second component, but we leave this for further work.

\section{Experiments}
\label{sec:experiments}

\begin{table*}[t]
    \captionsetup{font=small}
    \centering
    \caption{Automatic evaluation, large models with different codecs and tokenization. \textit{Seen speakers}: LibriTTS-R held-out set from train-clean-100. \textit{Unseen speakers}: 200 utterances from LibriTTS-R dev-clean. \textit{Out-of-domain}: 200 utterances from the VCTK dataset.}
    \vspace{-1pt}
    \begin{tabular}{l c | c c c | c c c | c c c}
    \toprule
      \multirow{2}{*}{Codec} & 
      \multirow{2}{*}{Tokens} &
      \multicolumn{3}{c}{Seen Speakers} &
      \multicolumn{3}{c}{Unseen Speakers} &
      \multicolumn{3}{c}{Out-of-Domain} \\
      & &
      WER,\%$\downarrow$&
      CER,\%$\downarrow$&
      SSIM$\uparrow$ &  
      WER,\%$\downarrow$&
      CER,\%$\downarrow$&
      SSIM$\uparrow$ &  
      WER,\%$\downarrow$&
      CER,\%$\downarrow$&
      SSIM$\uparrow$ \\
    % & 
    \midrule
      Ground Truth & –  & 2.70 & 0.93 & - & 2.42 & 0.87 & - & 1.02 & 0.44 & - \\
      \midrule
      DAC & BPE       & 6.66 & 3.15 & 0.903 & 7.52 & 3.77 & 0.876 & 3.86 & 2.13 & 0.762 \\
      NeMo-Codec & BPE       & 4.86 & 2.19 & \textbf{0.908} & 7.19 & 4.45 & \textbf{0.881} & 4.81 & 2.40 & 0.773 \\
      EnCodec & BPE       & 4.90 & 2.18 & 0.903 & 6.25 & 3.01 & 0.868 & 3.93 & 1.79 & 0.759 \\
      \midrule
      DAC & IPA       & 4.04 & 1.67 & 0.900 & 4.89 & 2.13 & 0.866 & 4.41 & 1.96 & \textbf{0.775} \\
      NeMo-Codec & IPA & \textbf{3.36} & \textbf{1.31} & 0.906 & \textbf{4.56} & \textbf{2.11} & 0.871 & \textbf{3.05} & \textbf{1.48} & 0.765 \\
      EnCodec & IPA   & 3.73 & 1.43 & 0.901 & 4.61 & 2.50 & 0.853 &  3.93 & 1.75 & 0.754 \\
   \bottomrule
    \end{tabular}
    \label{model-evaluation}
    \vspace{-7pt}
\end{table*}

\begin{table}[t]
  \captionsetup{font=small}
  \centering
  \caption{
    Impact of the model size. EnCodec codec, BPE tokens. \textit{Seen speakers}: LibriTTS-R held-out set from train-clean-100. \textit{Unseen speakers}: 200 utterances from dev-clean.
    }
  \vspace{-1pt}
  \resizebox{\columnwidth}{!}{
  \begin{tabular}{c c c | c c | c c}
  \toprule
      \multicolumn{3}{c}{Num Layers} & 
      \multicolumn{2}{c}{Seen Speakers} &
      \multicolumn{2}{c}{Unseen Speakers} \\
      Encoder &
      Precitor &
      RCH &
      WER,\% $\downarrow$ & SSIM$\uparrow$ &
      WER,\% $\downarrow$ & SSIM$\uparrow$ \\
  \midrule
     6 & 3 & 6 & 6.66 & 0.900 & 6.93 & 0.875 \\
     6 & 6 & 6 & 6.04 & 0.899 & 6.74 & 0.875 \\
     12 & 6 & 6 & 5.56 & 0.900 & 6.40 & 0.868 \\
     6 & 6 & 12 & 5.42 & 0.896 & 6.46 & \textbf{0.877} \\
     12 & 6 & 12 & \textbf{4.90} & \textbf{0.903} & \textbf{6.25} & 0.868 \\
    \bottomrule
    \end{tabular}
    }
    \label{ablation-studies}
\end{table}

\begin{table}[t]
  \captionsetup{font=small}
  \centering
  \caption{
    Challenging Texts Evaluation. Comparison of best TTS-Transducer models with external LLM-based TTS models. Naturalness MOS, 95\% confidence interval. Randomly selected male and female speakers, 92 utterances for each (184 total).}
  \vspace{-1pt}
  \resizebox{\columnwidth}{!}{
  \begin{tabular}{l c c c}
  \toprule
     Model & WER,\% $\downarrow$ & CER,\% $\downarrow$ & MOS$\uparrow$ \\
  \midrule
     Bark~\cite{BarkTTS} & 22.92 & 11.67 & 3.82 $\pm$ 0.04 \\
     VALL-E-X~\cite{zhang2023vallexspeakforeign} & 19.25 & 7.96 & 3.72 $\pm$ 0.04 \\
     SpeechT5~\cite{ao2021speecht5} & 16.24 & 6.00 & \textbf{3.84 $\pm$ 0.04} \\
     \midrule
     Ours, EnCodec, BPE & 17.28 & 5.66 & 3.83 $\pm$ 0.04 \\
     Ours, NeMo-Codec, BPE & 16.87 & 5.37 & 3.81 $\pm$ 0.04 \\
     \midrule
     Ours, EnCodec, IPA & 15.64 & 4.50 & 3.69 $\pm$ 0.04 \\
     Ours, NeMo-Codec, IPA & \textbf{13.83} & \textbf{3.94} & 3.82 $\pm$ 0.04 \\
    \bottomrule
    \end{tabular}
    }
    \label{challenging-texts}
    \vspace{-8pt}
\end{table}

\subsection{Datasets}
We use LibriTTS-R~\cite{koizumi2023LibriTTSR} dataset, an improved 24 kHz version of the LibriTTS ~\cite{zen2019LibriTTS}. The LibriTTS corpus contains a diverse set of speakers reading English audiobooks. We train the model on \texttt{train-clean-100}, \texttt{train-clean-360}, and \texttt{train-other-500} subsets of the LibriTTS-R. We set aside 1.15 hours of data from the \texttt{train-clean-100} to test the model on seen speakers. We filter the data by a maximum duration of 15 seconds, which results in 464 hours in the training dataset. To evaluate our system on \textit{unseen speakers}, we randomly choose 0.35 hours of data for unseen speakers with 39 unseen speakers from the \texttt{dev-clean} subset. Additionally, we evaluate our model in \textit{out-of-domain} conditions on a subset of 200 utterances (0.2 hours) from VCTK~\cite{yamagishi2019VCTK} to test the generalization abilities to multiple acoustic conditions. We use reference audios of length between 3 to 5 seconds.

\subsection{Model Details}
We use pretrained neural audio codecs: EnCodec~\cite{defossez2022encodec} (8 codebooks, 6 kbps), NeMo-Codec~\cite{langman2024spectralnemocodec} with RVQ (8 codebooks, 6.9 kbps), and Descript Audio Codec~\cite{kumar2023daccodec} model (9 codebooks, 8 kbps).
Our TTS-Transducer model has 12 Transformer layers in the encoder, 6 layers in the prediction network, and 12 layers in residual codebook head. The encoder and residual codebook head have 2 attention heads and a feed-forward dimension of 1536, with model dimension of 640 and 512, respectively. The prediction network Transformer-decoder uses 4 attention heads, with a model dimension of 512 and a feed-forward dimension of 2048. This results in \emph{199M} parameters for EnCodec and NeMo-Codec, and \emph{200M} parameters for the DAC model due to a larger embedding table since DAC codec uses 9 codebooks.
For the speaker embedding model, we use 1024 640-dimensional GST~\cite{wang2018globalstyletoken} learnable embeddings.

All TTS-Transducer models are trained in NeMo~\cite{kuchaiev2019nemo} toolkit with a global batch of 2048 for 200 epochs using 32 NVIDIA A100 GPUs. We use AdamW~\cite{loshchilov2017decoupledadamw} optimizer with cosine annealing scheduler~\cite{loshchilov2016sgdrcos} with 2000 warmup steps and a maximum learning rate of $1\mathrm{e}{-3}$.

During inference, we use nucleus sampling~\cite{holtzman2019nucleussampling} ($p = 0.95$) for predicting codes from the first codebook. The second component predicts remaining codebooks greedily.

\subsection{Results}
\label{sec:results}

We use automatic evaluation metrics -- Character Error Rate (CER), Word Error Rate (WER) and Speaker Similarity (SSIM) -- to compare our models with different audio codecs and tokenization. To test robustness, we transcribe generated audio using the pretrained Parakeet-CTC-1.1B\footnote{\label{parakeet-ctc}\url{https://hf.co/nvidia/parakeet-ctc-1.1b}} and compute CER and WER  with respect to the ground truth transcriptions. For SSIM, we compute cosine similarity between embeddings of generated speech and ground truth audio obtained from WavLM \cite{chen2022wavlm} model\footnote{\url{https://hf.co/microsoft/wavlm-base-plus-sv}}. The results are shown in Table~\ref{model-evaluation}. All the models provide intelligible speech with a relatively low word error rate. This shows that our model can produce robust speech and preserve the target speaker's acoustic qualities, regardless of the audio codec being used.
Most of the BPE-based models show best speaker similarity scores, but all models using IPA show significantly better intelligibility.

We perform ablation studies for BPE-based models varying the number of layers in encoder, predictor and residual codebook head. The results in Table~\ref{ablation-studies} show that all the components contribute to the intelligibility of the produced speech.

\textit{For comparison with external TTS systems}, along with automatic metrics, we use Mean Opinion Score (MOS) evaluations. Each audio was rated by at least 11 independent listeners on Amazon Mechanical Turk platform. Each rater was asked the question "How natural does the speech sound?" for a given audio on a scale of 1 to 5 with 1 point difference, 1 being totally unnatural and 5 extremely natural.
We choose following TTS models which use large-scale language modeling approach: VALL-E-X~\cite{wang2023valle}, Bark~\cite{BarkTTS} with EnCodec audio codec, and also SpeechT5~\cite{ao2021speecht5}\footnote{\url{https://hf.co/microsoft/speecht5_tts}}. For VALL-E-X, we use unofficial open-source implementation with checkpoint\footnote{\url{https://github.com/Plachtaa/VALL-E-X}} trained on 1739 hours of audio\footnote{\url{https://plachtaa.github.io/vallex/}}. 
Bark authors do not disclose\footnote{\url{https://github.com/suno-ai/bark/issues/2}} the amount of data but claim the work is comparable\footnote{\url{https://github.com/suno-ai/bark/issues/277}} with related AudioLM~\cite{borsos2023audiolm} and VALL-E~\cite{wang2023valle}, both using $\sim$60k hours of audio data. SpeechT5 uses LibriSpeech audio for pretraining (1k hours) and 400M text sentences from the text corpus and is finetuned on the LibriTTS dataset. 
As shown in Table~\ref{challenging-texts}, the proposed BPE-based TTS-Transducer outperforms Bark and public VALL-E-X systems in intelligibility and is comparable with best models in naturalness. The IPA-based model with NeMo codec is also comparable in naturalness but shows the best robustness across all the compared models.

\section{Conclusion}
\label{sec:conclusion}

We presented a novel TTS-Transducer system that predicts audio neural codec tokens directly from phonemes based on a neural transducer. It combines the strength of the neural transducer to learn monotonic alignment between text and audio and the effectiveness of neural audio codecs. The neural transducer predicts the first codebook. The remaining residual codes for the same frame are predicted by a separate block based on the learned alignment. Both components are optimized jointly. We demonstrated that our model can produce high-quality, reliable speech with popular audio codecs. Our experiments showed that the model achieves results comparable to SOTA TTS models in naturalness, and surpasses them in intelligibility on challenging texts without requiring large-scale pretraining.

\bibliographystyle{IEEEtran}
\bibliography{refs}

\end{document}